\documentclass{emulateapj}
\usepackage{natbib}
\usepackage{graphicx}

\usepackage{mathrsfs}

\begin{document}
\title{Predicted Number, Multiplicity, and Orbital Dynamics of {\it TESS} M Dwarf Exoplanets}

\author{Sarah~Ballard\altaffilmark{1,2}}

\altaffiltext{1}{MIT Kavli Institute for Astrophysics \& Space Research, Cambridge, MA 02139, USA; sarahba@mit.edu}
\altaffiltext{2}{MIT Torres Fellow for Exoplanetary science}


\begin{abstract}
We present a study of the M dwarf exoplanetary systems forthcoming from NASA's {\it TESS} mission. While the mission's footprint is too complex to be characterized by a single detection completeness, we extract an ensemble completeness function that recovers the M dwarf planet detections from previous work. We employ this completeness function, together with a dual-population planet occurrence model that includes compact multiple planetary systems, to infer anew the planet yield. We predict both the number of M dwarf planets likely from {\it TESS} and their system architectures. We report four main findings: first, that {\it TESS} will likely detect more planets orbiting M dwarfs that previously predicted. Around stars with spectral types between M1V--M4V, we predict {\it TESS} will find 990$\pm$350 planets orbiting 715$\pm$255 stars, a 1.5-fold increase over previous predictions. Secondly, {\it TESS} will find two or more transiting planets around 20\% of these host stars, a number similar to the multiplicity yield of NASA's {\it Kepler} mission. Thirdly, {\it TESS} light curves in which one or more planets are detected will often contain transits of additional planets below the detection threshold of {\it TESS}. Among a typical set of 200 {\it TESS} hosts to one or more detected planets, 116$\pm$28 transiting planets will be missed. Transit follow-up efforts with the photometric sensitivity to detect an Earth or larger around a mid-M dwarf, even with very modest period completeness, will readily result in additional planet discoveries. And fourth, the strong preference of {\it TESS} for systems of compact multiples indicates that {\it TESS} planets will be dynamically cooler on average than {\it Kepler} planets, with 90\% of {\it TESS} planets residing in orbits with $e<0.15$. 
\end{abstract}

\section{Introduction}

NASA's {\it TESS} Mission \citep{Ricker14} will furnish the vast majority of small, rocky planets for atmospheric study. A typical {\it TESS} target star receives 27 days of continuous observation, so the sensitivity of the mission strongly favors short periods \citep{Sullivan15}. A handful of transits of a small planet will be detectable over this duration only if those transits are individually large, which is why 75\% of small planets detected by {\it TESS} are expected to orbit M dwarfs \citep{Sullivan15}. In fact, it is likely that every small planet discovered by {\it TESS} to reside in its star's habitable zone will orbit an M dwarf \citep{Sullivan15}. Combining this fact with the favorable signal-to-noise ratio of a planetary transmission spectrum around a small star \citep{Tarter07}, M dwarfs will likely be the majority of sites for focused follow-up atmospheric study in the next decade with the James Webb Space Telescope ({\it JWST}, \citealt{Gardner06}). The forthcoming {\it TESS} sample of planets orbiting M dwarfs will likely contain many targets of the first biosignature searches. 

The ensemble of planets orbiting M dwarfs has come into focus from a combination of radial velocity, microlensing, high-contrast imaging, and transit surveys (\citealt{Bonfils13,Johnson10,Clanton16,Montet14,Bowler15,Dressing13, Dressing15,Morton14, Muirhead15}. For a detailed summary, see \citealt{Shields16}). In particular, the photometric sensitivity of NASA's {\it Kepler} mission illuminated the population of planets smaller than 4$R_{\oplus}$ in orbit around M dwarfs, showing that they are more common around late spectral types than around FGK dwarfs \citep{Howard12,Mulders15}. They are so common, in fact, that \cite{Morton14} found 2.00$\pm$0.45 planets per M dwarfs, and \cite{Dressing15} reported a similar value of 2.5$\pm$0.2 planets per star. 

\begin{figure*}
\includegraphics[width=\textwidth]{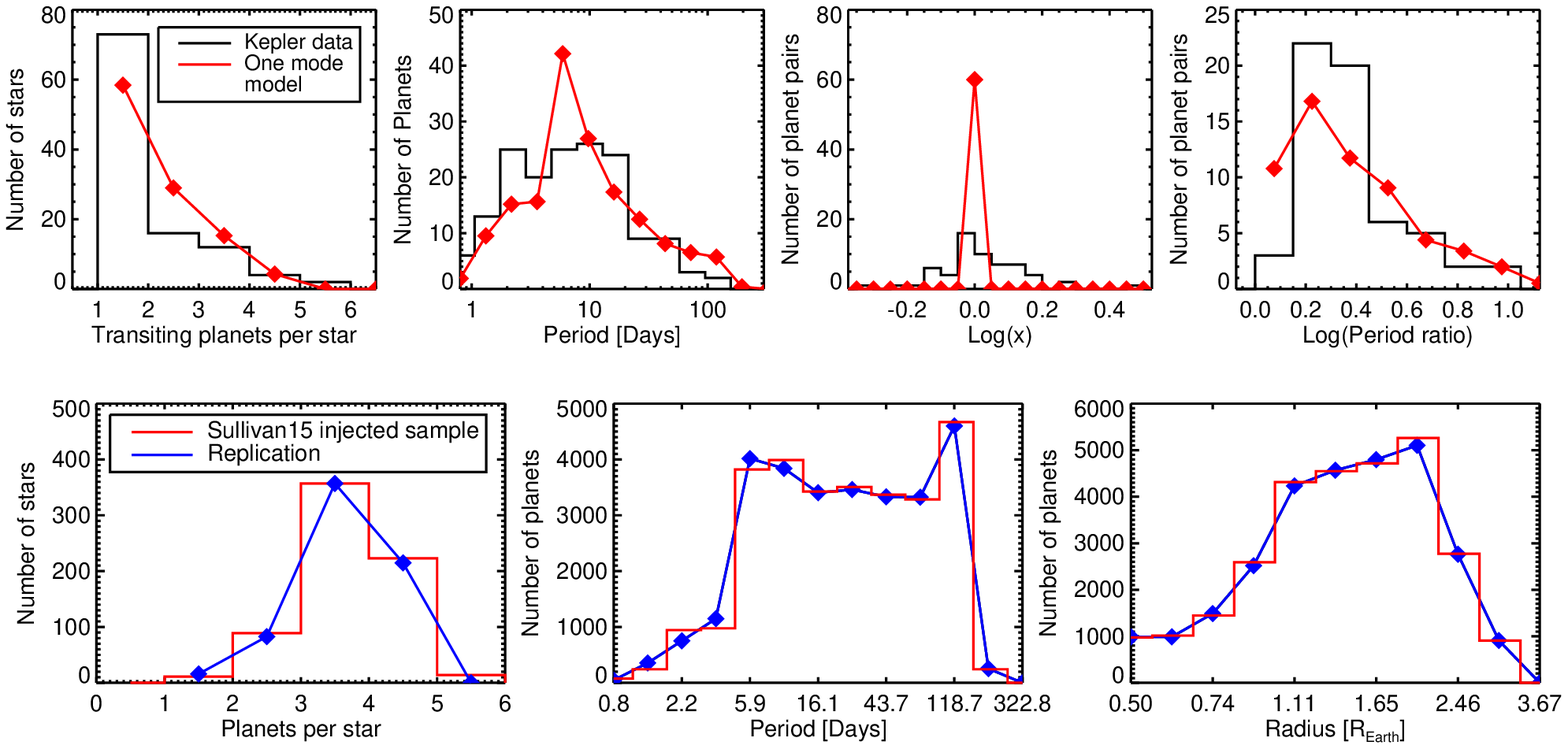}
\caption{{\it Top panels}: Observed {\it Kepler} distributions (black) of detected planets in number of transiting planets star$^{-1}$, period, transit duration ratio ($\xi$ defined in \citealt{Fabrycky12b}), and period ratio. Overplotted in red is the best one-mode planet occurrence model, with 2.5 planets per star drawn from \cite{Dressing15}. The delta function in $\xi$ in the third panel is due to uniforming applying an orbital mutual inclination of zero degrees. {\it Bottom panels}: Consistency in the underlying distributions of number of planets per star, periods, and radii of our one-mode model (red) to the values used in \cite{Sullivan15}.} 
\label{fig:replication}
\end{figure*}

Yet, M dwarf planetary systems resist a simple, one-population explanation. The top panels of Figure \ref{fig:replication} shows the result of using one mode of planet occurrence. Drawing 2--3 planets per star from the occurrence rate of \cite{Dressing15} furnishes only a fair fit to the {\it Kepler} properties of detected planets orbiting M dwarfs \citep{Ballard16}. One explanation is that the model of 2--3 planets per star, with the underlying period and radius distribution in \cite{Dressing15}, is in fact an average of two very different types of planetary systems. Observations of orbital eccentricity and spin-orbit alignment indicate that the systems with one transiting planet are dynamically distinct from those with two or more transiting planets. Planets in multiple-planet systems reside in more circular orbits \citep{Xie16}, and are more aligned with the spins of their host stars \citep{Morton14b}. The number of transiting planets per star from {\it Kepler} also indicates two populations with different dynamical properties, one with at least 5 planets coplanar to within 2$^{\circ}$, and other with 1--2 planet at larger orbital inclinations with respect to one another \citep{Ballard16}. This two-population model removes the discrepancy in the top left panel of Figure \ref{fig:replication}, in which the number of systems with only one transiting planet is underestimated, and the number of systems with two transiting planets is overestimated in equal measure. The two-population model also furnishes a better fit to other observables, like period, period ratio, and transit duration ratio \citep{Dawson16,Moriarty16}.  While the so-called ``{\it Kepler} dichotomy" \citep{Lissauer11b} explanation is not definitive, nor the only one \citep{Gaidos16, Bovaird17}, we employ it here as a useful phenomenological descriptor of M dwarf planetary systems.

This two-mode model is also consistent with the independent measurement on the rate of ``compact multiples": these are systems with at least two planets with orbital periods less than 10 days. \cite{Muirhead15} showed that at least 20\% of M dwarfs host a compact multiple system, and that fraction increases as stellar temperatures decreases. Within the two-population framework, these compact multiples are recognizable as the systems with $>5$ planets per star interior to 200 days. As we described above, compact multiples need to be included into order to reproduce the {\it Kepler} yield \citep{Ballard16}. Similarly, their inclusion should also result in a more realistic prediction of the {\it TESS} yield. 

A sophisticated study of the likely {\it TESS} planet yield across the FGKM spectral types by \cite{Sullivan15} incorporated the complicated {\it TESS} footprint, its instrumental limitations, the range of noise budgets within the surveyed stellar population, and false positive likelihoods. However, it employed the one-mode model of planet occurrence. The study undertaken by \cite{Muirhead17} specific to the {\it TESS} M dwarfs accounted for additional complexities in the sample, particularly with respect to selecting exposure times. That study also assumed a single mode of planet occurrence. We hypothesize that an occurrence model that includes two distinct types of planetary systems, rather than one model that averages the the two, will result in a predicted {\it TESS} sample that will be different in important ways:

\begin{itemize}
\item The sample will contain more planet detections.
\item It will find that {\it TESS} will detect two or more transiting planets around a substantial number of stars 
\item It will find that the {\it TESS} lightcurves with a detected planet will very often contain transits of additional planets lurking below the noise. 
\end{itemize}


 We do not aim to replicate the \cite{Sullivan15} machinery in its complexity: because of that study, we already have an excellent understanding of how {\it TESS} will respond to incoming photons. Rather, we propose to extend the analysis for a different planet occurrence rate. In order to accomplish this, we need to extract the {\it TESS} completeness function for planets orbiting M dwarfs as a function of radius and period. This function is not included in the \cite{Sullivan15} study, but is readily derivable from it. Before we expand upon the result of \cite{Sullivan15}, we must demonstrate that we can replicate it, by showing that this completeness function correctly recovering the \cite{Sullivan15} detections from their sample of injected planets . As for any survey, the {\it TESS} completeness function we will extract will be an average of the individual completenesses for each star observed by the mission. 
 
 
 With the {\it TESS} completeness in hand, we can apply it to a different sample of injected planets. We organize this study as follows: in Section 2, we describe our analysis, including the generation of synthetic planetary systems ($\S$2.1) and how we create a mixture model of planetary systems ($\S$2.2). Section $\S$2.3 describes the extraction of the M dwarf completeness function for {\it TESS}, and Section $\S$2.4 describes how we apply it to the mixture model. Section 3 contains the results of this exercise. We enumerate the following goals for this study, which are addressed in the indicated sections. 
 
 \begin{enumerate}
     \item Re-predict the number of planet detections among M dwarfs observed by {\it TESS} ($\S$3.1)
     \item Determine how often {\it TESS} will detect a single transiting planet, and how often it will detect two or more planets transiting the same star ($\S$3.2)
     \item Determine which additional planets, if any, will transit known {\it TESS} planet host stars but elude detection in {\it TESS} light curves ($\S$3.3)
    \item Predict the fraction of {\it TESS} detected systems that will have the ``compact multiple" architecture, as compared to the underlying rate in nature ($\S$3.4)
     \item Predict the eccentricity distribution of the detected {\it TESS} planets, and compare it to that of {\it Kepler} M dwarf planets ($\S$3.5)
     \item Approximate the number of planets {\it TESS} will detect that will exhibit transit-timing variations (TTV), using the rate of TTV occurrence measured by {\it Kepler} ($\S$3.6)
    \item Make a prediction for the bulk densities of planets detected by {\it TESS}, from planet formation theory. Compare these densities to the densities inferred for the {\it Kepler} planets. ($\S$3.7)
 \end{enumerate}
 
 In Section 4, we summarize our findings and conclude. 
 
\section{Analysis}
\label{sec:analysis}

\subsection{Generating Planetary Systems}
\label{sec:generate}

To generate a realistic synthetic samples of planetary systems, we take the following steps. We draw periods and radii for each mock planetary system from the empirical distribution of \cite{Dressing15}. We then employ the distributions of \cite{Limbach15} to assign eccentricity. We assign planetary masses with the relations of \cite{Zeng17} for $R<1.5R_{\oplus}$ and \cite{Wolfgang16} for $R>1.5R_{\oplus}$. \cite{Rogers15} identified the cutoff between majority of rocky planets and a majority icy/gaseous planets at 1.5$R_{\oplus}$, but these two relations also naturally overlap at 1.5 $R_{\oplus}$. We assess the stability of the system by ensuring that planets satisfy the criteron defined in \cite{Fabrycky12b}:

\begin{equation}
\Delta \equiv (a_{2}-a_{1})/R_{H_{1,2}} > 2 \sqrt{3},
\label{eq:delta}
\end{equation}

where the mutual Hill Radius $R_{H_{1,2}}$ is defined by

\begin{equation}
R_{H_{1,2}}=\Big[\frac{M_{1}+M_{2}}{3M_{\star}} \Big] ^{1/3}\frac{(a_{1}+a_{2})}{2}.
\end{equation}

This criterion is applicable for circular orbits. For eccentric orbits, we calculate the periapse and apoapse separation from the host star for each planet. We assume the orbits are stable if Equation 1 holds for the apoapse distance of the inner planet and the periapse of the outer planet. For generating synthetic {\it TESS} planetary systems, we employ a stellar mass of 0.4$M_{\oplus}$. 

We then assign a boolean transit timing variation (TTV) flag to each transiting planet. \cite{Xie14} showed that planets drawn from multi-transiting systems are likelier to exhibit TTVs, with that likeliness increasing as the number of transiting planets increases. We assign TTV probability per planet from that work, as defined by their ``Case 3" (the most generous TTV occurrence rate): 3.5\% per planets in singly-transiting systems, 7\% for planets in doubly-transiting systems, 8\% for planets in triply-transiting systems, and 10.4\% for planets in systems with 4 or more transiting planets. 

Finally, we calculate and record an independent density for each planet using only its mutual Hill spacing from neighboring planets. \cite{Dawson16} predicted a theoretical relationship between these parameters: $\rho$ [g/cm$^{3}$]=$(\Delta / 22)^{6}$, where $\Delta$ is defined in Equation \ref{eq:delta}. The scaling of density with mutual Hill separation provides a natural explanation, among others, for the fluffier nature of planets whose masses were measured with transit-timing variations \citep{Wolfgang16, Mills17}.  

\subsection{Generating Mixture Models}
\label{sec:mixture}

In the simplified ``{\it Kepler} dichotomy" model, stars host one of two distinct types of planetary systems. \cite{Ballard16} showed that the {\it Kepler} M dwarf planets are well-described by one population of stars hosting flat and manifold systems of planets (with number of planets per star $N$ at least 5, and orbital mutual inclinations $\sigma$ between 1 and 3$^{\circ}$), with the other hosting 1 planet or 2 planets with high mutual inclination ($>$8$^{\circ}$). Throughout this work, we refer to the former type of planetary system as ``Population 1" or more descriptively as a ``compact multiple". That work investigated the mixture specifically among {\it detected planet hosts}: in reality, the former type of planetary system is overly represented among detected planet hosts. This is because typical short periods within the multiple systems make it likelier that at least one planet will transit. The degree of this overrepresentation in both {\it Kepler} and {\it TESS} is discussed in Section \ref{sec:occurrence}. 

We define $N$ as the number of planets per star and $\sigma$ as the width of the Rayleigh distribution from which we draw their mutual inclinations. For a set of \{$N$,$\sigma$\},  we generate 10$^{4}$ planetary systems using the criteria established in Section \ref{sec:generate}, employing the posterior distributions on this quantities from \cite{Ballard16}. We then determine, based upon the assumption of random alignments for each planetary system with the line of sight, which planets transit their host star. We consider non-integer $N$ as follows. If a ``typical" planetary system defined by $N$ and $\sigma$ hosts 3.5 planets, then 50\% of stars host 3 planets (with eccentricities drawn from the CDF for 3-planet systems) and 50\% host 4 (with eccentricities drawn from the CDF for 4-planet systems). 

The Fraction $f$ of stars in Population 1 in \cite{Ballard16} is the fraction of {\it transiting} systems in Population 1, not the fraction of stars. For each $f$, we now calculate an $f_{\star}$, the fraction of stars in Population 1 necessary to recover a contribution $f$ to the total number of {\it transiting} systems. For mixture models defined by the set \{$N_{1},\sigma_{1},N_{2},\sigma_{2},f_{\star}$\}, we randomly select 10$^{4}\cdot f_{\star}$ stars populated by \{$N_{1},\sigma_{1}$\} and 10$^{4}\cdot$(1-$f_{\star}$) stars populated by \{$N_{2},\sigma_{2}$\}. We then draw properties from the transiting planets among this final set of 10$^{4}$ stars. Of course, the vast majority of these stars host zero transiting planets. 

\subsection{Extracting the {\it TESS} Completeness Function}
\label{sec:completeness}

To determine what a transit mission will detect, it is useful to know how often any given transiting planet will be detected, typically as a function of radius and orbital period. This quantity is the ``completeness," and it is in principle distinct for every star the mission surveys. For example, when measuring {\it Kepler}'s completeness to planets transiting FGK dwarfs, \cite{Christiansen16} performed an injection-and-recovery exercise for each of 198,154 target stars. In absence of real light curves, \cite{Sullivan15} used a signal-to-noise criterion to evaluate whether injected planets would be ``detected" by {\it TESS}. The stare-and-step observation strategy of {\it TESS} means that most stars in the mission footprint receive 27 days of continuous photometry. However, overlap between observing fields results in some stars residing for up to a year in the field of view. \cite{Sullivan15} study incorporated the full complexity of the {\it TESS} footprint and its stellar sample. The completeness of the survey to planets of a given size and period is not included in the study, but is derivable from it. 

\begin{figure}[htb]
\centering
  \begin{tabular}{@{}c@{}}
    \includegraphics[width=.40\textwidth]{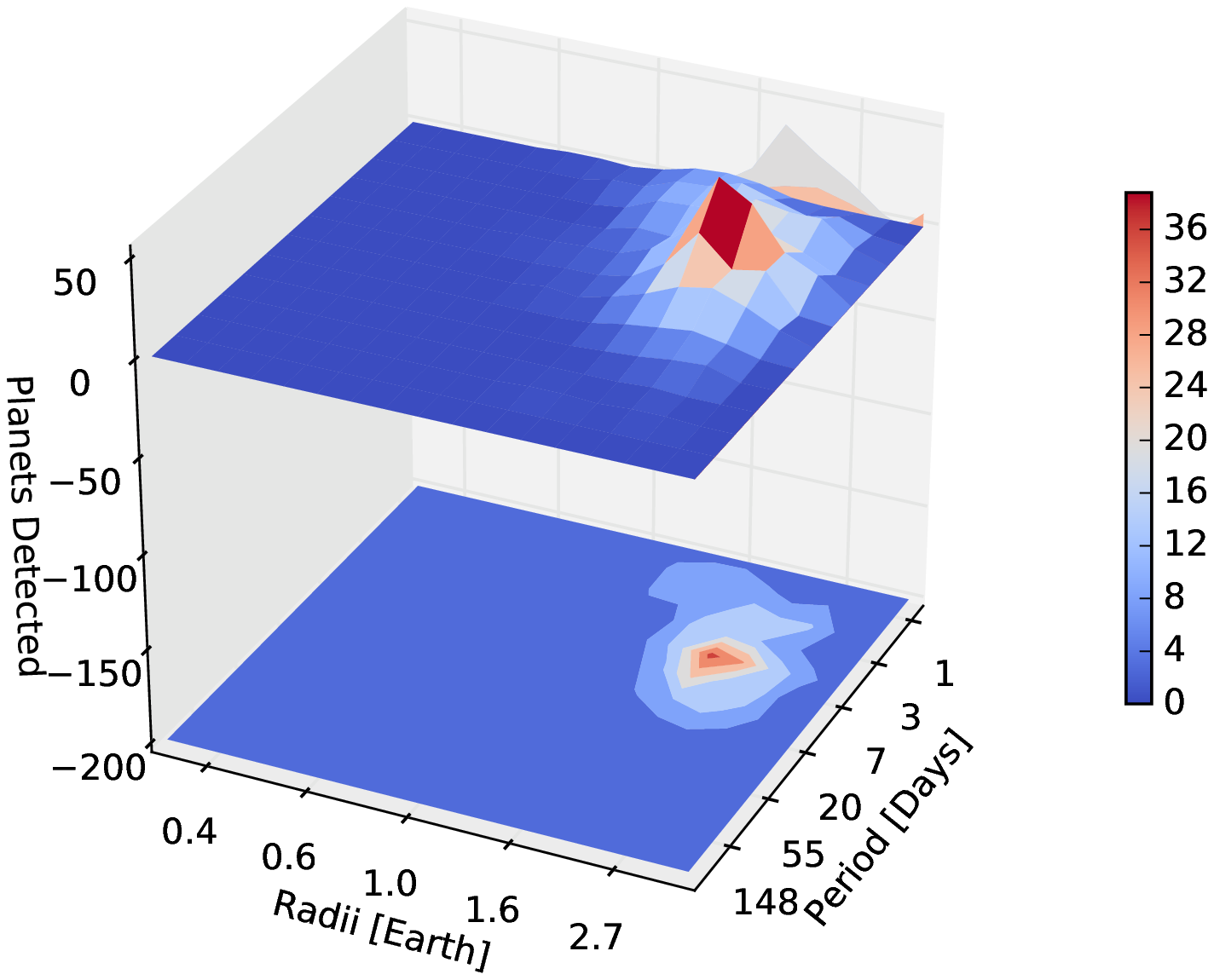} \\
    \includegraphics[width=.40\textwidth]{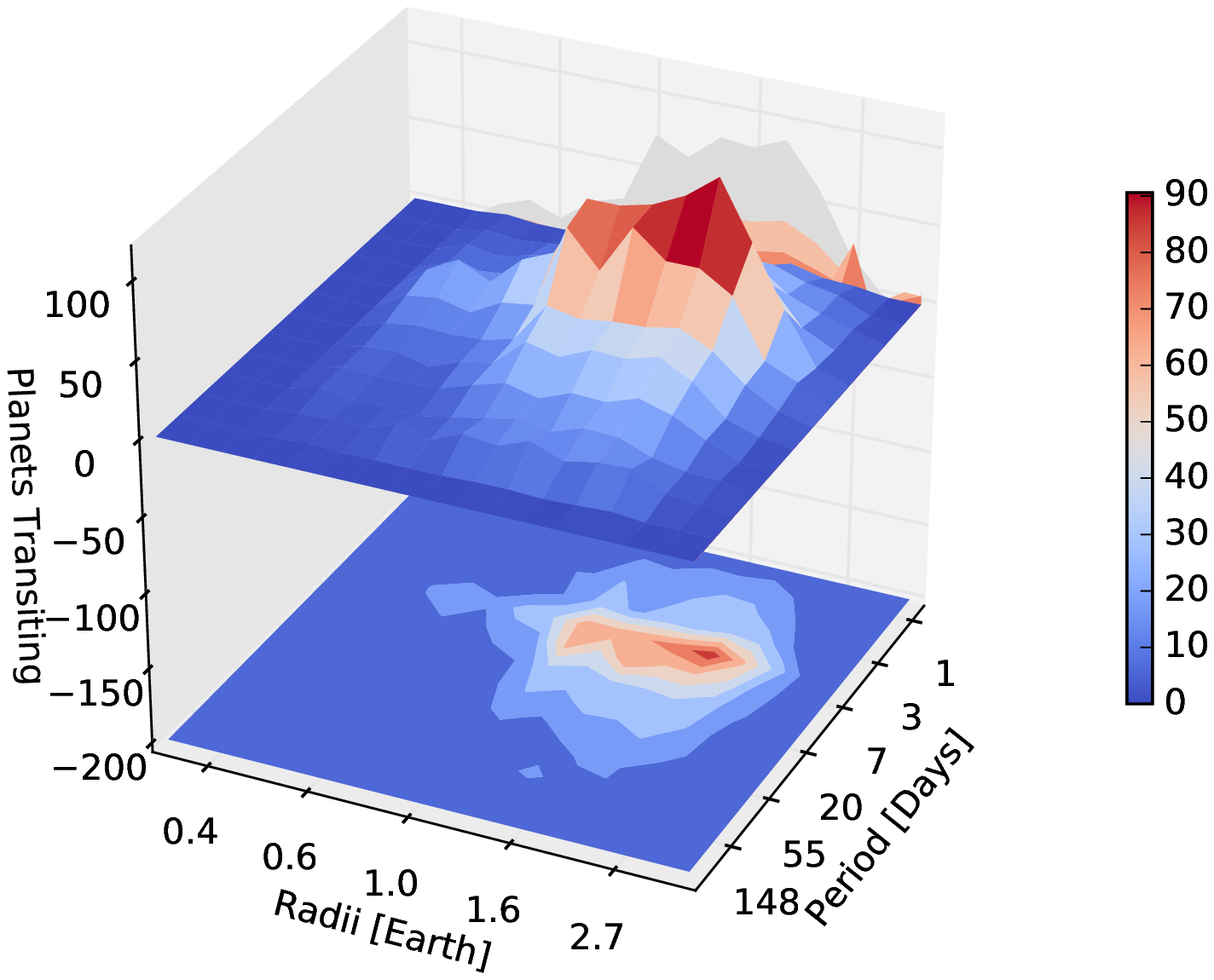} \\
    \includegraphics[width=.40\textwidth]{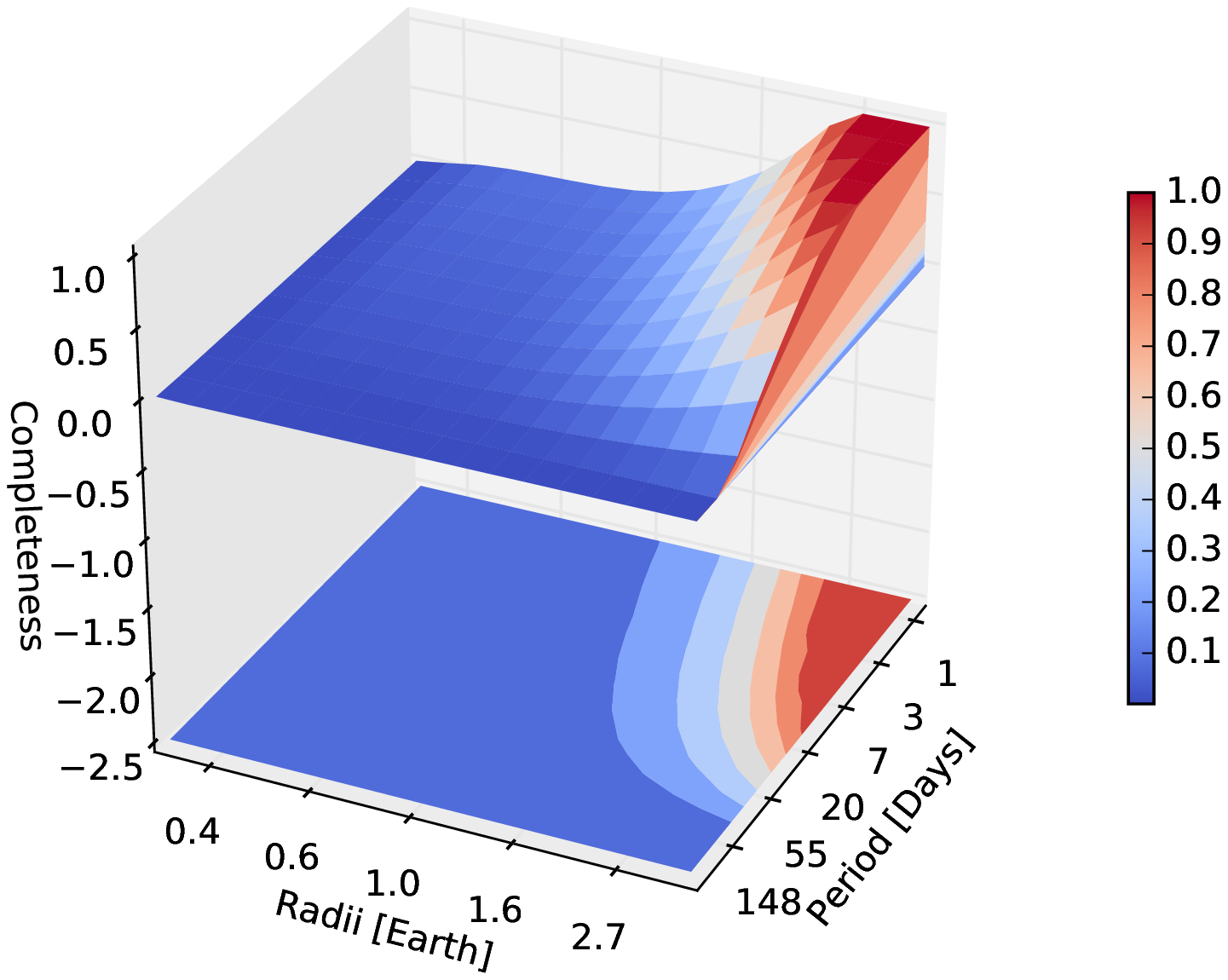} \\
  \end{tabular}
  \caption{{\it Top panel:} Detected planets ${D_{i,j}}$ from \cite{Sullivan15}. {\it Middle panel:} Injected transiting planets sample generated using the criteria from same, ${N_{i,j}}$. {\it Bottom panel:} Typical polynomial completeness function evaluated at each bin. ${C_{i,j}}$.}
\label{fig:completeness_surface}
\end{figure}

We first create a sample of injected planets, using the criteria described in \cite{Sullivan15}. This process is similar to the one described in Section \ref{sec:generate} for a single mode of planet occurrence, except that (1) they assign more than one planet to a given star with independent probability, rather than assigning the number of planets per star {\it a priori} and (2) they assume a mutual inclination between orbits of zero. In the bottom panels of Figure \ref{fig:replication}, we show consistency between the sample we generated from the stated criteria and the one employed in that study. We aim now to find the completeness function that winnows this injected sample to the detected sample reported by \cite{Sullivan15}. That sample was generated at fixed resolution in both log(period) and log(radius) (inherited from \citealt{Howard12}, \citealt{Dressing13}, and others), with approximate spacing of 1 dex between adjacent log(period) bins, and 0.2 dex between adjacent log(radius) bins. Because we cannot expect to extract information at a higher resolution than these values, we adopt their spacing to create a grid in log(period) and log(radius) space. We treat each individual bin at this resolution as a bucket that holds an integer number of planet detections. We calculate the number of detected planets from \cite{Sullivan15} in each bin. We call this ${D_{i,j}}$, where $i$ is the index of the period bin and $j$ the index of the radius bin. In practice, the index $i$ spans periods from 0.8 to 320 days, in 13 regular log intervals of 1 dex, and the index $j$ spans radii from 0.3 to 4 $R_{\oplus}$, in 17 regular log intervals of 0.2 dex.  Similarly, we calculate the number of {\it injected} planets in each bin, ${N_{i,j}}$. The actual surfaces of both ${D_{i,j}}$ and ${N_{i,j}}$ are shown in Figure \ref{fig:completeness_surface}. We multiply the number of injected planets in each bin by the completeness corresponding to that bin to produce the model number of detections in each bin, ${\mu_{i,j}}.$ 

We experimented with various functional forms for the completeness, which we call $C$. We adopt a smooth, analytic function for $C$, which we evaluate at the same resolution in log(period) and log(radius) to produce the completeness of each bin ${C_{i,j}}$. Trial verions of $C$ included a single-to-noise scaling, as well as simple power laws in log period or radius. Neither of the two functions for completeness, when applied to the injected planets, correctly approximated the number of detected planets: for example, while the predicted number of short-period planets might match, long-period planets would be strongly underestimated. We elected to use a polynomial in log(radius) and log(period) for $C$, with some constraints. 

First, we require the completeness to be separable in period and radius (that is, $C(P,R_{p})=C(P)\cdot C(R_{p})$. We require that it be bounded between 0 and 1. And we require it to be monotonic (increasing with radius and decreasing with period). Any monotonic polynomial can be modeled as a series of polynomials of the form \cite{Hawkins94}:

\begin{equation}
G(x)=a_{4}+a_{1}\int_{0}^{x} \prod_{i=1}^{K} \big[1+2a_{2,i}t+(a_{2,i}^{2}+a_{3,i}^{2})t^{2} \big] dt,
\end{equation}

where all coefficients $a$ are all real numbers. Finding the optimal coefficients for a necessarily monotonic fit is a nonlinear problem. We first simplify it by using only the first term in the series, so that the completeness in both period and in radius have the form:

\begin{equation}
C(x)=a_{4}+a_{1}x+a_{1}a_{2}x^{2}+\frac{a_{1}}{3}(a_{2}^{2}+a_{3}^{2})x^{3}.
\end{equation}

This results in a total of 8 free parameters: 4 coefficients for the completeness with orbital period $C(P)$, and four for the completeness with planet radius $C(R_{p})$. There does exist a linear algebra solution for finding [$a_{1},a_{2},a_{3},a_{4}$] for each polynomial, a modified version of singular value decomposition adapted for monotonic polynomials that relies upon Lagrange multipliers \citep{Hawkins94,Murray13}. The \texttt{MonoPoly} package in R \citep{Monopoly16} implements those tools, which we used as a first estimate for the coefficients ${a}$: 4 for $C(P)$ and 4 for  $C(R_{p})$. 

Poisson counting statistics describe
integer numbers of transiting planets, so we evaluate the likelihood of ${a}$ with a Poisson likelihood function.  This likelihood is conditioned on the ``observed" number of planets detected in that bin $D_{i,j}$. The model number of planet detected in each bin  $\mu_{i,j}(a)$ is dependent upon ${a}$ as follows:

\begin{equation}
\mu_{i,j}(a)=N_{i,j}\cdot C_{i,j}(a),
\end{equation}

where ${N_{i,j}}$ is the number of injected planets in that bin and ${C_{i,j}(a)}$ is the completeness polynomial with coefficients ${a}$, evaluated at each bin. Figure \ref{fig:completeness_surface} shows the surfaces of $D_{i,j}$, $N_{i,j}$, and $C_{i,j}(a)$ for a set of ${a}$. In the top panel, we show the detected planets published by \cite{Sullivan15}, ${D_{i,j}}$. The second panel shows our replication of the sample of the injected planets ${N_{i,j}}$. And the third panel shows a typical completeness ${C_{i,j}}$.

The final Poisson likelihood function is defined as follows:

\begin{equation}
\mathscr{L} \propto \prod_{i}\prod_{j}\frac{\mu_{i,j}(a)^{D_{i,j}}e^{-\mu_{i,j}(a)}}{D_{i,j}!}.
\label{eq:likelihood}
\end{equation}

\begin{figure*}
\includegraphics[width=\textwidth]{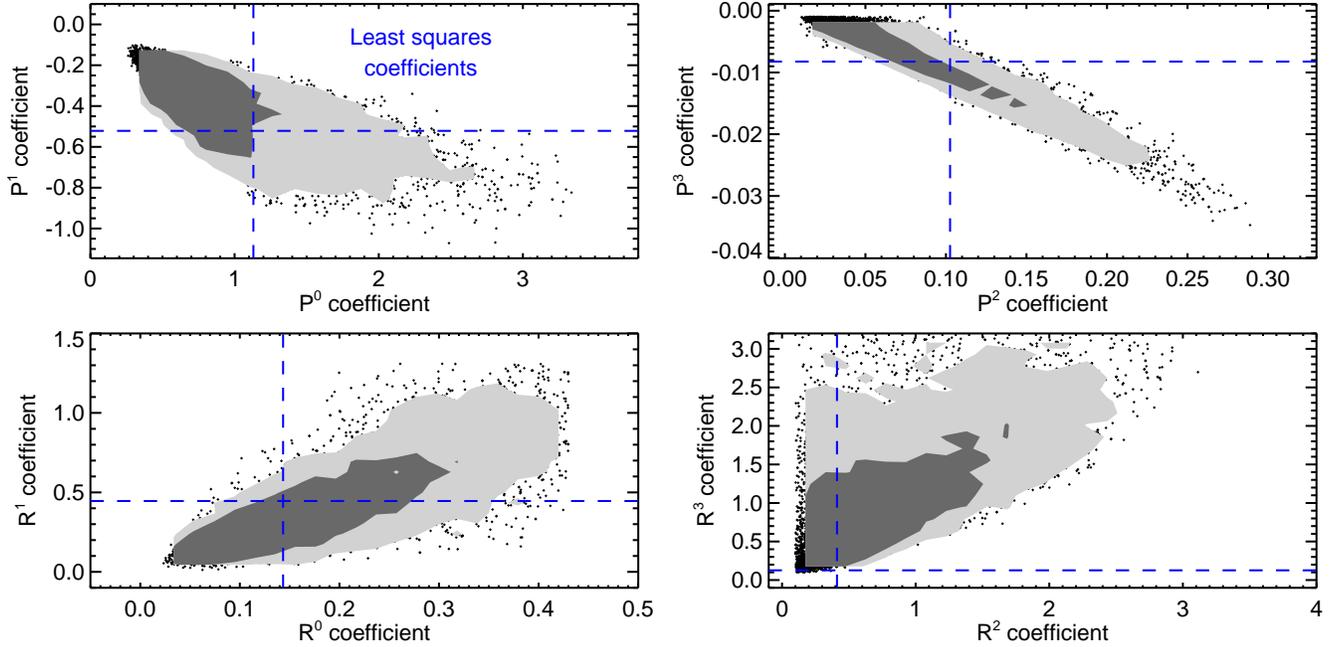}
\caption{Posterior distributions for completeness function coefficients. Top two panels show the four coefficients for (log) period completeness (the coefficient for $P^{2}$ is indicated by the axis labeled $P^{2}$), while bottom two panels show coefficients for (log) radius completeness. Blue dotted lines show least-squares solution coefficients for a monotonic polynomial fit.}
\label{fig:coefficients}
\end{figure*}

We employ the Bayesian sampler
MultiNest (\citealt{Feroz08,Feroz09,Feroz13}, with Python implementation by \citealt{Buchner14}) to
evaluate these likelihoods and posterior distributions. In
practice, MultiNest calculates the log of the likelihood 
defined in Equation \ref{eq:likelihood}. We use uniform priors for each of the polynomial coefficients, allowing them to vary to within 200\% of the least-squares value. We enforce a monotonically decreasing polynomial in log period and a monotonically increasing polynomial in log radius, by setting the log likelihood to an arbitrarily low value otherwise (-$10^{-30}$, in our case, in comparison to a typical log likelihood of -300). Figure \ref{fig:coefficients} shows the posterior distributions for all 8 coefficients, with the least-squares values overplotted in blue. In 6 of 8 cases, the least-squares value lies within the 1$\sigma$ contour of the posterior distribution, and within 2$\sigma$ in all cases. 

\begin{figure*}
\includegraphics[width=\textwidth]{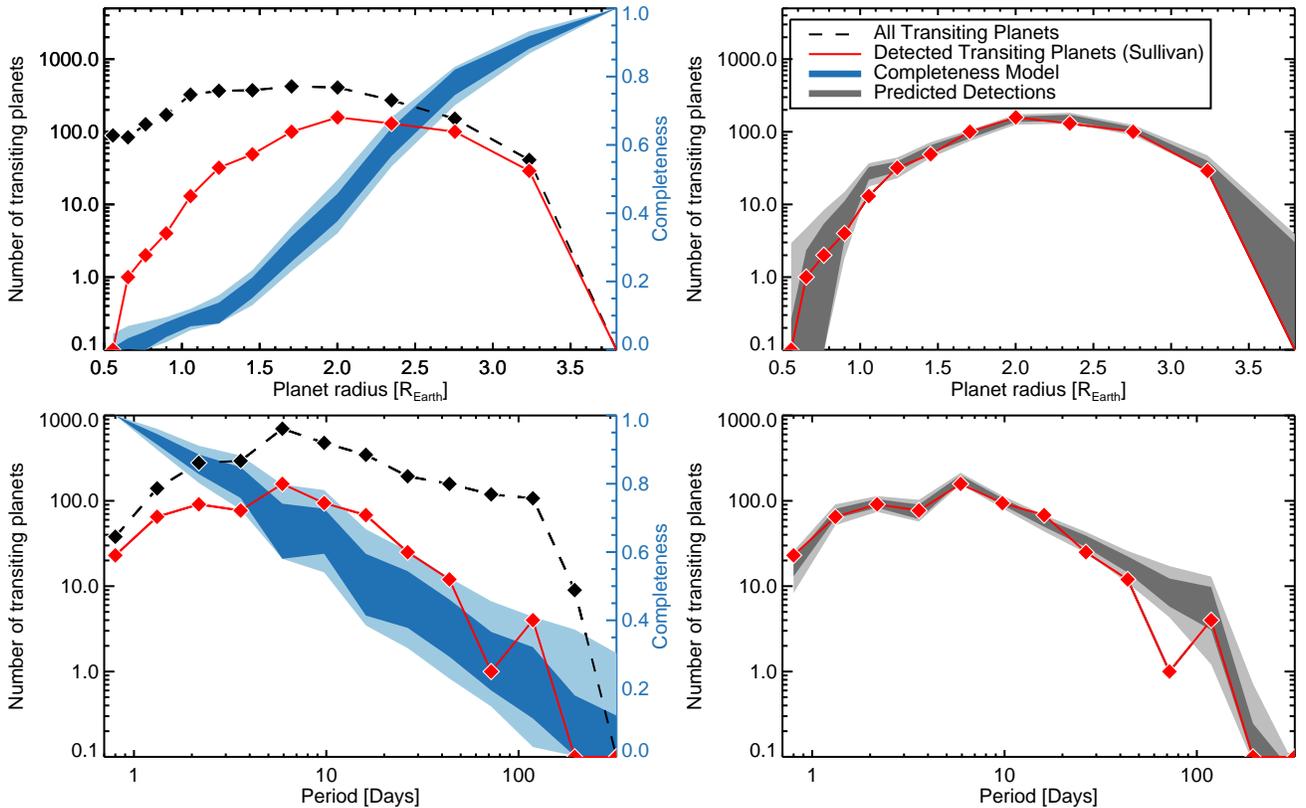}
\caption{{\it Left panels:} Injected transiting planets ${N}$ (black) and detected planets ${D}$ (red) from \cite{Sullivan15}, as function of radius and period. Model completeness functions in radius and period are overplotted in dark (1$\sigma$) and light (2$\sigma$) blue: the right axis corresponds to completeness. {\it Right panels:} The population of detected \cite{Sullivan15} planets, now with models $\mu$ for predicted planet detections overplotted in gray.}
\label{fig:completeness_results}
\end{figure*}

Figure \ref{fig:completeness_results} summarizes the results of the completeness fit. The injected planets ${N_{i,j}}$ are shown in black as a function of radius (top panel) and period (bottom panel). The detected planets ${D_{i,j}}$ are shown in red. Blue shows the completeness functions that best recovers ${D_{i,j}}$ from ${N_{i,j}}$, which we drawing from the posterior distribution in the coefficients ${a}$. The right-hand panels show the 1 and 2$\sigma$ confidence intervals on the model number of detected planets ${\mu_{i,j}}$ in grey. We verify that the extracted completeness is successful at recovering the detected planets from the injected planets of \cite{Sullivan15}. We note the large uncertainty of the completeness function at long periods: at 100 days, for example, completenesses of both 40\% and 0\% are consistent at 2$\sigma$ confidence. This is due to the inherent Poisson noisiness of only a few ($<10$ planets) detections with which to constrain the completeness.

\subsection{Applying Completeness to Occurrence Mixture Model}

With a {\it TESS} completeness function in hand, we can apply it to a new sample of simulated transiting planets, this time employing the mixture model in \cite{Ballard16}. As described in Section \ref{sec:generate}, we use the posteriors in the number of planets in both systems, $N_{1}$ and $N_{2}$, their average mutual inclinations $\sigma_{1}$ and $\sigma_{2}$, and the fraction of host stars in the first population $f$ directly from that work.
The completeness coefficient posteriors in Figure \ref{fig:coefficients} show that the coefficients are highly correlated. We cannot draw independently from their posterior distributions anew to sample the completeness. Rather, we save the completeness surface at each iteration of the MCMC chain.   

For each transiting planetary system, we draw randomly from the sample of completeness surfaces ${C_{i,j}}$. For each individual planet's period and radius, we evaluate the detection likelihood from the completeness value corresponding to that bin. We take one additional step to enforce consistency for planets orbiting the same star. A joint random draw of detection probabilities for multiple planets can occasionally result in the nonsensical scenario of less-likely planets being detected, while more ``detectable" planets are missed. We take as an example a system of two transiting planets: Planet 1 with a period and radius assigning it a 50\% detection probability and Planet 2 with a radius and period assigning it a 5\% detection probability. Out of 200 draws for a set of two random numbers between 0 and 1, there will be 5 instances in which Planet 2 is detected and Planet 1 is missed. This is a sensible scenario for an ensemble of stars, but for planets orbiting the same star, we consider the SNR ratios of the two planets (assume the same noise budget for both), and if the missed planet has a higher SNR than the detected planet, we switch the former to ``detected." 

We record the properties of each ``detected" transiting planet, as well as its provenance (whether from a dynamically hot or cool configuration). For the sake of comparison, we repeat the exercise with the completeness function of {\it Kepler} \citep{Dressing15}, so that we we can directly compare {\it Kepler} observables to those predicted for {\it TESS}. To generate synthetic {\it Kepler} systems, we employ a stellar mass $M_{\star}=0.50M_{\odot}$ as compared to $M_{\star}=0.40M_{\odot}$ for {\it TESS} to reflect a typical M dwarf from both surveys (we note that altering the central mass by 20\% results in only small changes to the resulting observables).

\section{Results}

We revisit the goals enumerated in Section 1. 

\subsection{Summary of Planet Detections}
\label{sec:results}
\begin{figure*}[htb]
\centering
\begin{tabular}{ll}
    \includegraphics[width=3.0in]{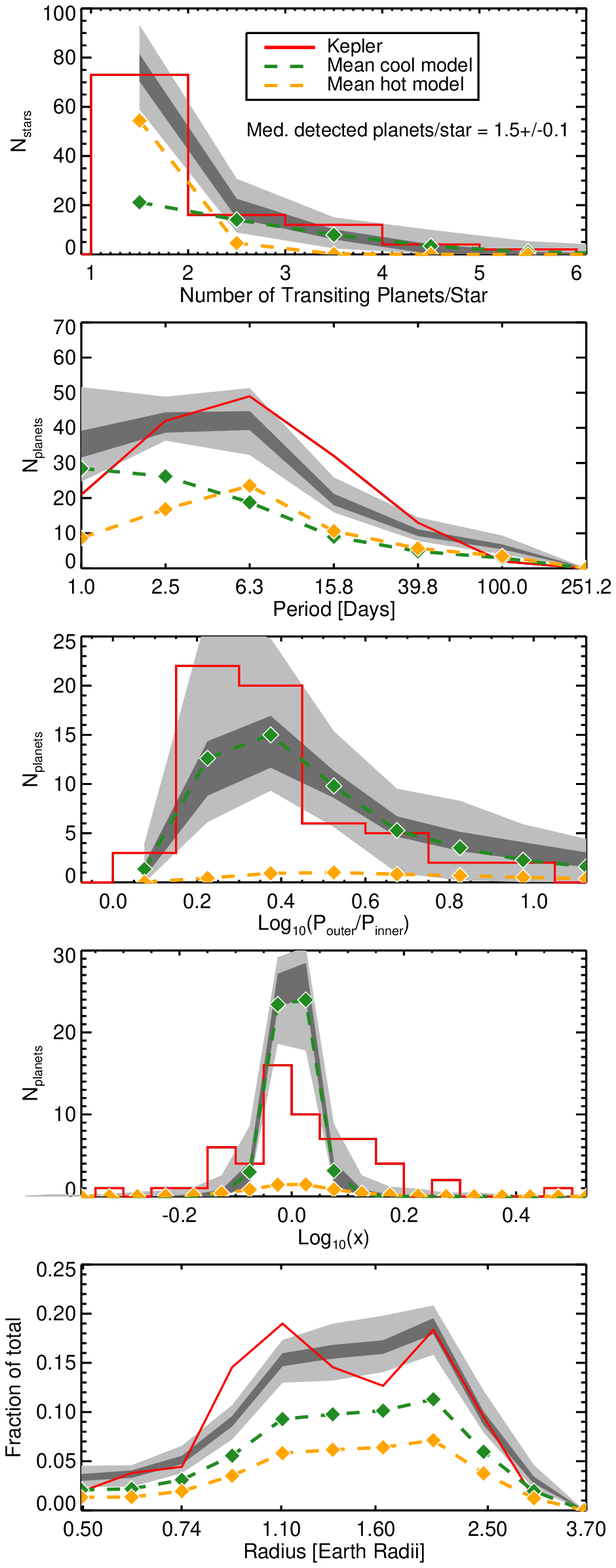} & \includegraphics[width=3.0in]{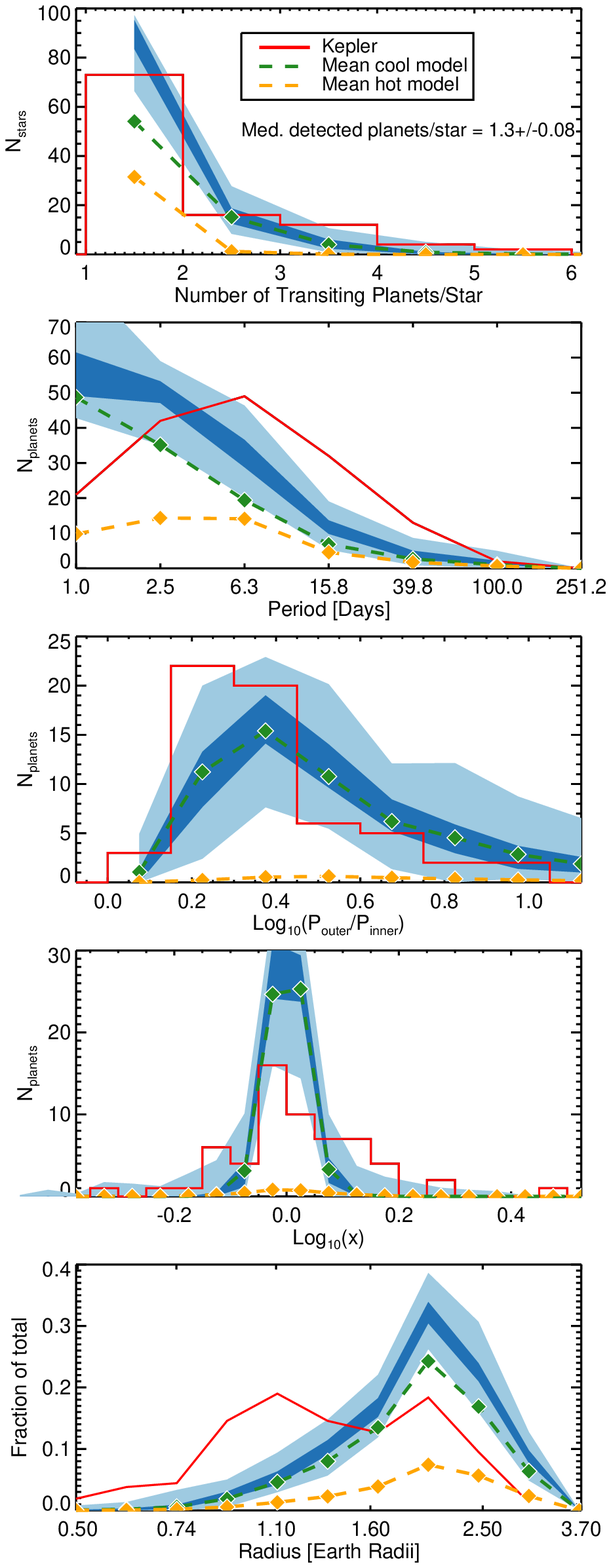}\\
  \end{tabular}
  \caption[width=1.0\textwidth]{Resulting posterior distributions predicted for the {\it Kepler} mission at left (grey) and {\it TESS} mission at right. In order from top to bottom: number of detected transiting planets per star, periods of detected planets, period ratio between adjacent observed transiting planets, velocity-normalized transit duration ratio between adjacent planets, and planetary radius. All distributions have been normalized to compare to the shape of the actual observed {\it Kepler} distributions in red.}
\label{fig:completeness1}  
\end{figure*}

\begin{figure*}[htb]
\centering
\begin{tabular}{ll}
    \includegraphics[width=3in]{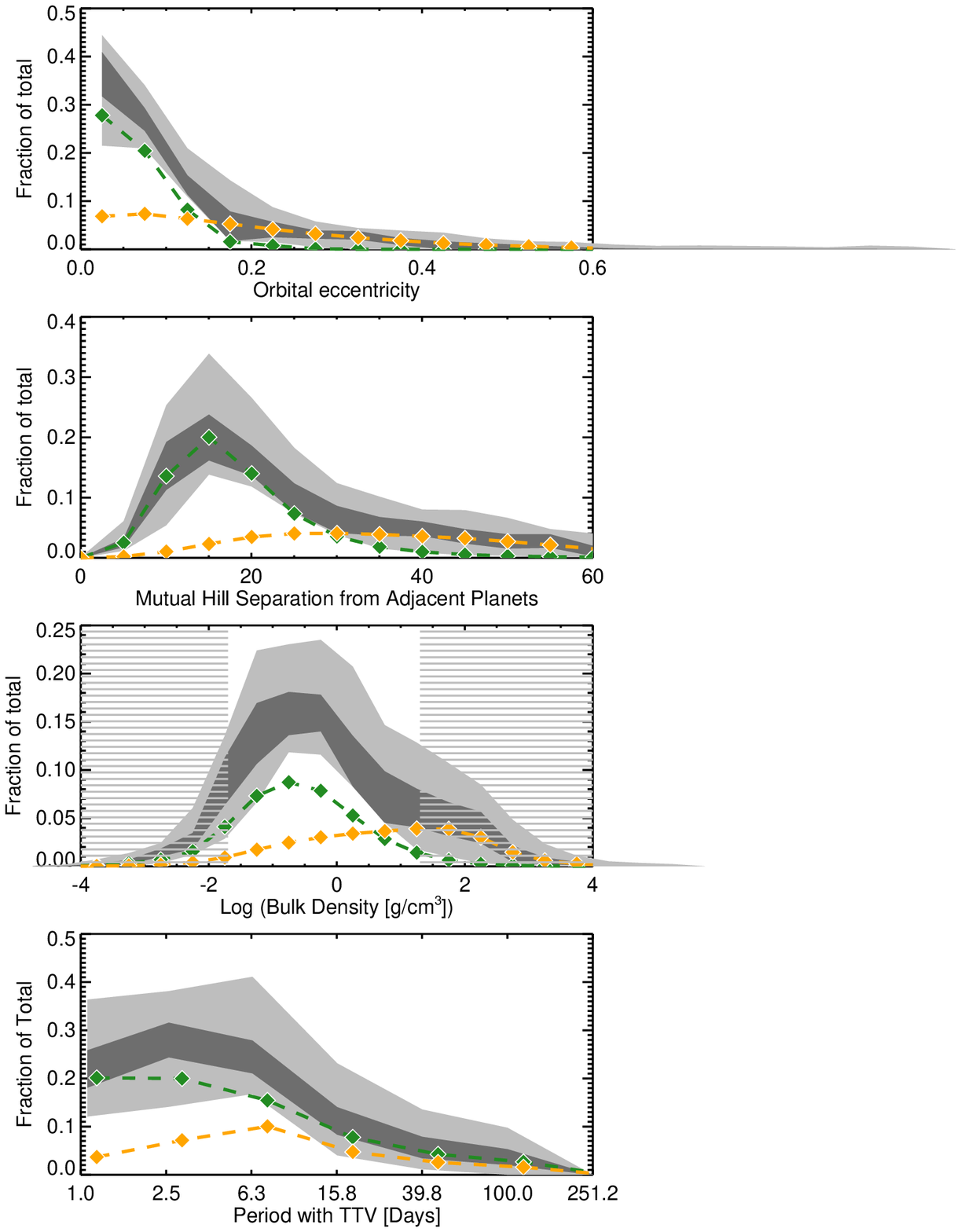} & \includegraphics[width=3in]{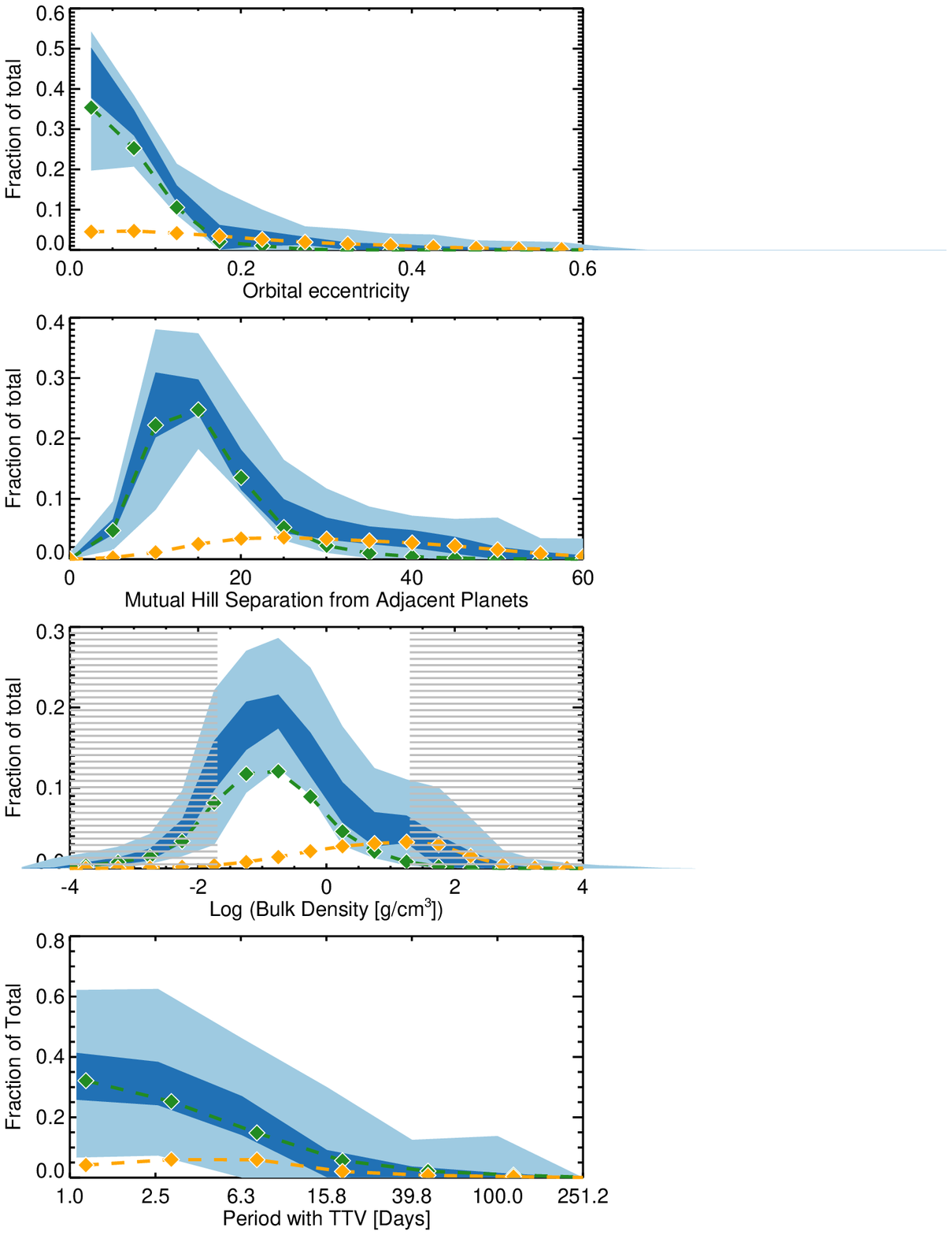}\\
  \end{tabular}
  \caption{Resulting posterior distributions predicted for the {\it Kepler} mission at left (grey) and {\it TESS} mission at right. In order from top to bottom: orbital eccentricity of detected planets, mutual Hill spacing from neighboring planets, predicted density from Hill spacing per \cite{Dawson16}, and periods of planets showing transit timing variations}.
\label{fig:completeness2}
\end{figure*}

In Figures \ref{fig:completeness1} and \ref{fig:completeness2}, we show the resulting distribution of properties for the M dwarfs observed by the {\it Kepler} (in grey) and {\it TESS} missions (in blue).  The distributions shown in Figure \ref{fig:completeness1} are normalized to compare directly to the {\it Kepler} yield, shown in red, while the distributions in Figure \ref{fig:completeness2} are normalized to one. The transit duration ratio, here denoted as $\xi$, is the one defined by \cite{Fabrycky12a}. For each parameter, we show the mean contribution to the total distribution from the dynamically cooler Population 1 (green) and the dynamically hotter Population 2 (orange). The first immediately noticeable difference is in the period and radius distributions, where the effects of the {\it TESS} completeness are clear. {\it TESS} will skew heavily toward detecting larger planets than {\it Kepler} and at shorter orbital periods. We note that the mutual Hill spacing distribution shown in the second panel of Figure \ref{fig:completeness2} is the true mutual spacing, not the (wider) spacing that would be measured between only {\it detected} planets. 

In addition to the shapes of these distributions, it's useful to note the raw number of expected host stars and host planets. From the posterior distributions in the modeled number of detections $\mu{i,j}$, we estimate that the {\it TESS} mission will find 990$\pm$350 planets orbiting 715$\pm$255 early-to-mid M dwarf host stars.   Unsurprisingly, given their strong representation among detected systems, the largest contribution to the uncertainty budget on the number of planets is the uncertainty on the fraction of planetary systems in compact multiples (see Section \ref{sec:occurrence}). 

We also investigate the subset of small, cool planets likely to be prioritized for follow-up with {\it JWST}. We define ``small" here to be radii $<2R_{\oplus}$ and ``cool" to be periods $20<P<40$ days (approximating the habitable zone of an M4V dwarf). Among the 990$\pm$350 planets detected by {\it TESS}, 15$\pm$7 meet this criteria. Critically for transit follow-up, an additional 39$\pm$20 planets in this radius and period range are undetected, but orbit stars for which {\it TESS} detected {\it another} planet. For the likeliest rocky planets with radii $<1.25R_{\oplus}$ in the same period range, {\it TESS} will detect 2$^{+2}_{-1}$ (consistent with the 3 planets with radii $<1.5R_{\oplus}$ and with orbital periods $>20$ days in the sample published in \citealt{Sullivan15}). In even starker contrast with slightly larger planets, 19$\pm$12 such planets will orbit known {\it TESS} hosts, but elude detection by {\it TESS} proper. In the hypothetical situation where each known {\it TESS} M dwarf host received 40 days of uninterrupted follow-up observation, the yield in newly uncovered temperate Earths would be triple or more that of {\it TESS} itself. It stands to reason that follow-up efforts with even moderate sensitivity at longer periods will uncover one or two of these, comparable to the number found in the mission data alone. We describe follow-up implications in greater detail in Section \ref{sec:followup}. 

\subsection{Multiple-transiting systems} 
Among these estimated 715$\pm$255 M dwarf planet hosts identified by {\it TESS}, the mission will detect at least two planets around 140$\pm$70 stars. The approximate 20\% contribution of multis to the host star budget is similar to {\it Kepler} (see top panel of Figure \ref{fig:completeness1}). Even with the steepness of its completeness function with period, we predict the mission will detect 32$\pm$19 systems with 3 more or transiting planets: a number that makes intuitive sense given the fact that 20\% of mid-M dwarfs host 2 or more planets interior to 10 days, and the average {\it TESS} star will receive 27 days of coverage. 

\begin{figure*}
\includegraphics[width=1.0\textwidth]{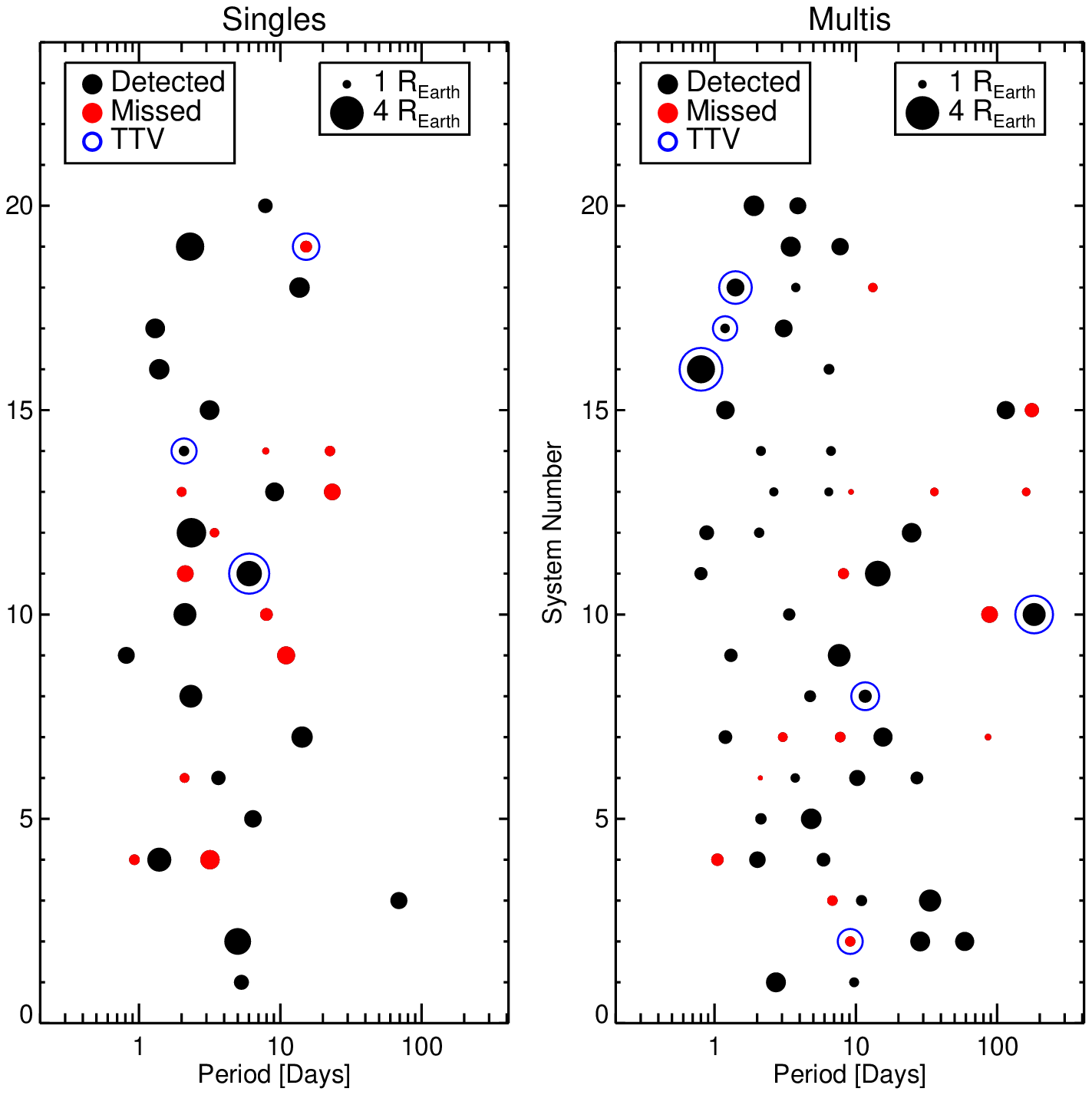}
\caption{{\it Left panel:} A representative sample of M dwarf planetary systems in which TESS detected a single transiting planet. We depict only the transiting planets orbiting each star. Black indicates TESS-detected planets, while red indicates planets that were missed. Planet radii is show by the relative sizes of circles, while planets exhibiting TTV are ringed in blue. {\it Left panel:} A simple sample for systems in which TESS detected two or more transiting planets.}
\label{fig:sample_systems}
\end{figure*}

Figure \ref{fig:sample_systems} shows a representative sample of {\it TESS} singles and multis, with a random selection of 20 systems from each population. Black circles, scaled to planet size, depict detections, where red circles are missed planets. The steep {\it TESS} radius completeness is especially evident visually here. We have indicated with blue circles the planets that exhibit TTV, assigned from the \cite{Xie14} occurrence rates as described in Section \ref{sec:generate}. The much higher rate of TTV among multi-transiting systems (even if only one planet was detected) is visually apparent: indeed, with a 3\% occurrence of TTV among singly transiting systems, none ought to appear in such a small representative sample. We note for clarity that we have shown 20 representative singly-transiting and 20 representative multiply-transiting systems as seen by {\it TESS} for a sense of their architectures. However, 50:50 is not representative of their relative contributions to the total {\it TESS} yield as we describe above.   

\subsection{Implications for transit follow-up} 
\label{sec:followup}
\begin{figure}
\includegraphics[width=.50\textwidth]{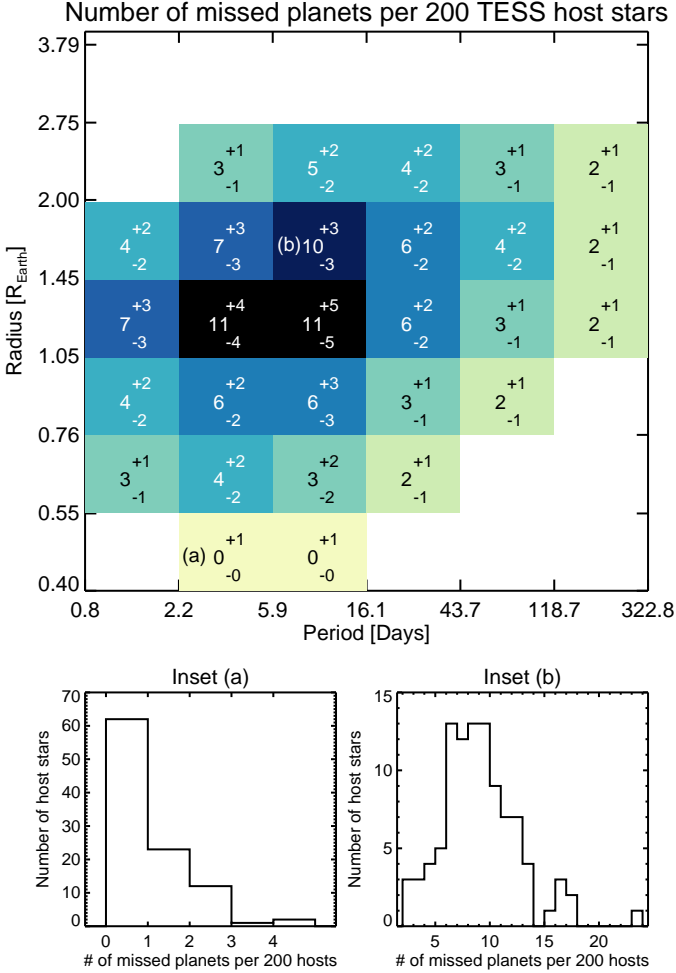}
\caption{{\it Top panel}: For every 200 {\it TESS} host stars, the number of transiting planets below {\it TESS}' detection threshold as a function of planetary radius and period.{\it Bottom panels}: The Poisson distributions in number of missed planets per 200 host stars for two example bins, indicted by (a) and (b) in the top panel.} 
\label{fig:missed_all}
\end{figure}

The best-fit ensemble completeness function for TESS in the top left panel of Figure \ref{fig:completeness_results} has a critical implication specific to transit follow-up. As a whole, {\it TESS} will see a 1.5 $R_{\oplus}$ planet 20\% of the time (though the exact completeness depends on the period of the planet as well). Comparing this modest likelihood with the occurrence rates of both \cite{Morton14} and \cite{Dressing15} (both of which show planets peaking in occurrence at 1-1.5 $R_{\oplus}$), it's clear that the majority of planets orbiting M dwarfs will be missed. Yet,  the majority of stars around which {\it TESS} finds a planet will host a compact multiple system (primarily because of the steep period completeness function). For systems of 3 or more transiting planets, 40\% of the time TESS will detect only one planet, typically the very largest. This means that missed planets orbiting known {\it TESS} hosts will be remarkably common. We quantify this result in Figure \ref{fig:missed_all}, showing the number of missed planets per 200 {\it TESS} host stars (that is, stars for which TESS detected one or more planets). Among 200 {\it TESS} host stars, typically 250 planets will be detectable in the mission light curves themselves. But on average, half that number lurk below the mission sensitivity: 116$^{+28}_{-28}$ planets per 200 host stars. This is also visibly apparent in Figure \ref{fig:sample_systems}, where red (missed) planets are common among their detected (black) neighbors. 

Therefore, follow-up efforts sensitive to planets $<1.5R_{\oplus}$, even those with very modest period completeness, will readily find additional planets. For example, a hypothetical survey of 200 TESS hosts, sensitive to 1$R_{\oplus}$ planets, with 100\% completeness out to only 2.2 days, will find an average of 11 additional planets (at least 6, and as many as 16, within the 68\% confidence interval). Especially promising for transit follow-up efforts: there are enough missed planets that surveys with 25\% completeness at 40 days can expect to find at least one rocky (1--1.5$R_{\oplus}$) planet in its habitable zone (for an M4V dwarf). 

We point out the subtle but important distinction between {\it number of planets missed} per 200 host stars and the {\it number of those stars} that host at least one missed planet. Among 200 hosts to {\it at least one} transiting planet, on average 120 host only that transiting planet. Among the remaining 80 hosts: 50 host one missed planet, 20 host 2 missed planets, and 10 host 3 or more. The odds are statistically distinct for hosts to one {\it TESS}-detected planet versus multiple detected planets, and also hosts to planets with detected transit-timing variations. For the sake of illustration, among a sample of 200 systems where {\it TESS} detected at least 2 transiting planets, now on average 95 host at least one additional transiting planet. And among a sample of 200 systems with a {\it TESS}-detected planet that also exhibits TTV, the odds are yet more favorable: 112 hosts out of 200 would have additional unseen planets among them.

\subsection{Underlying Occurrence Rate}
\label{sec:occurrence}
We have employed the posterior distribution in $f$ from \cite{Ballard16}, where $f$ in that work is the fraction of {\it transiting} systems in the compact-multiple configuration. However, because selection bias favors the detection of these systems, they are over-represented among the {\it Kepler} host star sample compared to their true underlying fraction among stars. We aim to test for consistency with occurrence rate for compact multiples orbiting M dwarfs derived by \cite{Muirhead15}, though we approach the problem in different ways. Firstly, the definition of ``compact multiple" in \cite{Muirhead15} is 2 or more planets interior to a 10 day orbit. Practically speaking, if planets are spaced equally in log semi-major axis, on average 5 $M_{\oplus}$, and dynamically stable, this corresponds to systems with 7 or more planets interior to 200 days. This value is safely within the posterior distribution of number of planets per star $N_{1}$ found by \cite{Ballard16}: 40\% of the distribution lies at 7 planets/star or greater. Secondly, \cite{Muirhead15} employed inverse-detection-efficiency machinery and compared to the number of stars hosting 2 or more planets compared to the number hosting no planets. In comparison, in \cite{Ballard16}, we ignored entirely systems hosting no transiting planets. We employed forward modeling to compare models to a different observable altogether: the shape of the distribution in the number of transiting planets per star. We invert the posterior in $f$ to $f_{\star}$ as follows.   

\begin{figure}
\includegraphics[width=.50\textwidth]{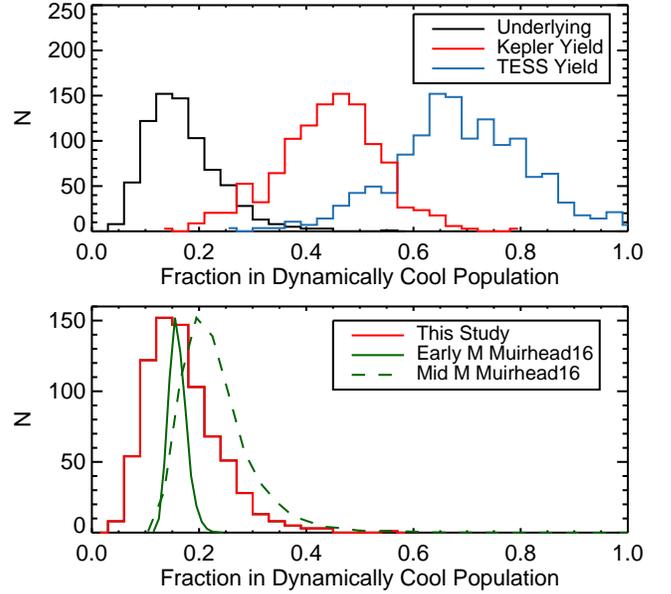}
\caption{{\it Top panel:} Fractions of systems in the compact-multiple configuration. {\it Bottom panel:} Same population of detected \cite{Sullivan15} planets, now with models for predicted planet detections overplotted in gray.}
\label{fig:compact_multi}
\end{figure}

For each 10$^{4}$ planetary systems we generate from the posterior on $f$ (described in Section \ref{sec:mixture}, we solve empirically for the fraction of stars $f_{\star}$ in Population 1. We make the assumption that every M dwarf in the sample hosts a planetary system of some kind, whether in Populations 1 or 2, so that the fractions sum to one. This assumption brings consistency between the mean number of planets per star of 2.0-2.5 determined by \cite{Morton14} and \cite{Dressing15}, and our planetary mixture model in which some stars host 5 planets and others host 1, as we described in \cite{Ballard16}. We record this fraction $f_{\star}$ at each step of the MCMC chain. The lower panel of Figure \ref{fig:compact_multi} shows the resulting distribution in $f_{\star}$, as compared to results from \cite{Muirhead15}. While compact multiples make up 45$\pm10$\% of transiting systems found by {\it Kepler}, they are only 15$\pm5$\% of all planetary systems orbiting early M dwarfs. This is consistent with the 15.9$\pm$1.5\% found by \cite{Muirhead15} for early M dwarfs. 

We compare in Figure \ref{fig:compact_multi} the resulting distributions in $f_{\star}$ between the {\it Kepler} and {\it TESS} missions. The selection bias that favored the detection of compact multiples from the {\it Kepler} Mission is still greater for NASA's {\it TESS} mission. We show in the top panel of Figure \ref{fig:compact_multi} the fraction of compact multis within the sample of planet hosts for both {\it Kepler} and {\it TESS}. Now the fraction $f$ is 68$\pm$12\%, showing that the steep period completeness for {\it TESS} will likely result in 5x the rate of compact multiples among {\it TESS hosts} than the underlying rate in nature. 

\subsection{Implications for ensemble eccentricity}

Orbital eccentricity in M dwarf planetary systems has complicated implications for habitability. For the smallest stars, even a modest eccentricity can induce a sterilizing ``runaway greenhouse" effect \citep{Barnes13}. On the other hand, modest eccentricity may be sufficient to induce plate tectonics in the absence of radiogenic heating \cite{Jackson08a}. The eccentricity of a planet also shapes how we interpret its atmospheric signature (for a detailed summary, see \citealt{Shields16}). 

\begin{figure}
\includegraphics[width=.50\textwidth]{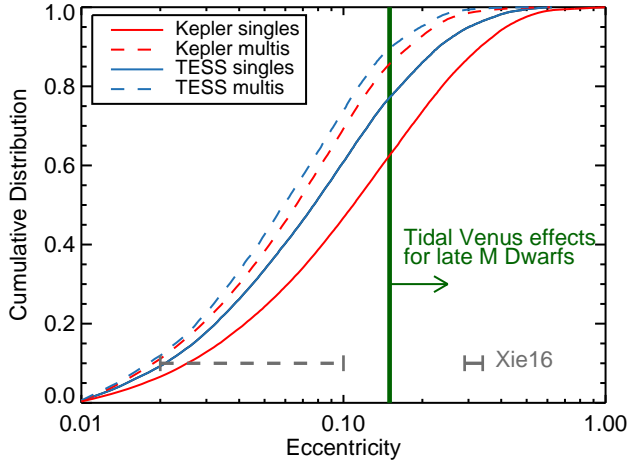}
\caption{The predicted cumulative eccentricity distribution for both {\it TESS} (blue) and {\it Kepler} (red) missions, where planets in a system with a single detected planet (solid line) are shown separately from planets in systems with multiple detected planets (dashed line). We have overplotted in gray the empirical eccentricites measured for {\it Kepler}'s singles (dashed) and multis (solids) from \cite{Xie16} for comparison.}
\label{fig:eccentricities}
\end{figure}

{\it TESS}' strong selection bias for shorter periods favors the discovery of compact and generally dynamically cooler systems generally, which we quantify in the previous section. This is particularly true for the multiple-planet systems uncovered by TESS, whose membership is almost certainly in this population.  Figure \ref{fig:eccentricities} summarizes this result, showing the cumulative eccentricity distributions for both the single and multiple transiting systems. A comparison between the {\it TESS} and {\it Kepler} distribution shows the predicted lower eccentricity for {\it TESS} planetary systems on average. This effect is strongest for the multi-transit systems from {\it TESS}, for which 80\% of planets have orbital eccentricities less than 0.1. We overplot the empirical result of \cite{Xie16} for the {\it Kepler} singles and multiples, measured from photometry. The eccentricities inferred for {\it Kepler} singles from this study are lower than the average measurement from \cite{Xie16}. However, we consider here the subset of M dwarfs rather than the full {\it Kepler} sample examined by \cite{Xie16}.  

We overplot, for the sake of comparison, an eccentricity associated with runaway tidal heating on late M dwarfs from \cite{Barnes13}. There exists a range of cutoff eccentricities for this effect, depending upon bulk planet composition, atmospheric composition, and assumptions about exactly how the dynamical heating occurs. For this reason, the cutoff shown here is illustrative rather than definitive. We have depicted the cutoff eccentricity of 0.15 for the sake of comparison with the cumulative eccentricity distributions. For higher orbital eccentricities, planets 1 $M_{\oplus}$ and larger in the habitable zone of 0.25$M_{\odot}$ stars are predicted to experience a runaway Venus effect \citep{Barnes13}. We note that {\it TESS} planets are safer from this effect on average, with the planets in multi-planet systems safest (with only 10\% possessing orbital eccentricities greater than 0.15).  

\subsection{Transit-timing Variations}

\begin{figure}[htb]
\centering
  \begin{tabular}{@{}c@{}}
    \includegraphics[width=.50\textwidth]{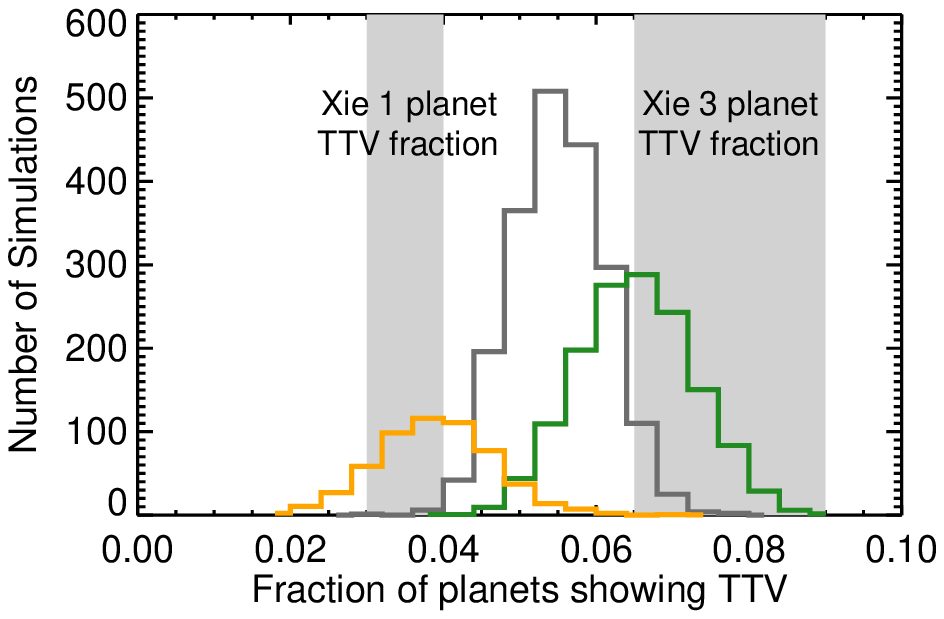} \\
    \includegraphics[width=.50\textwidth]{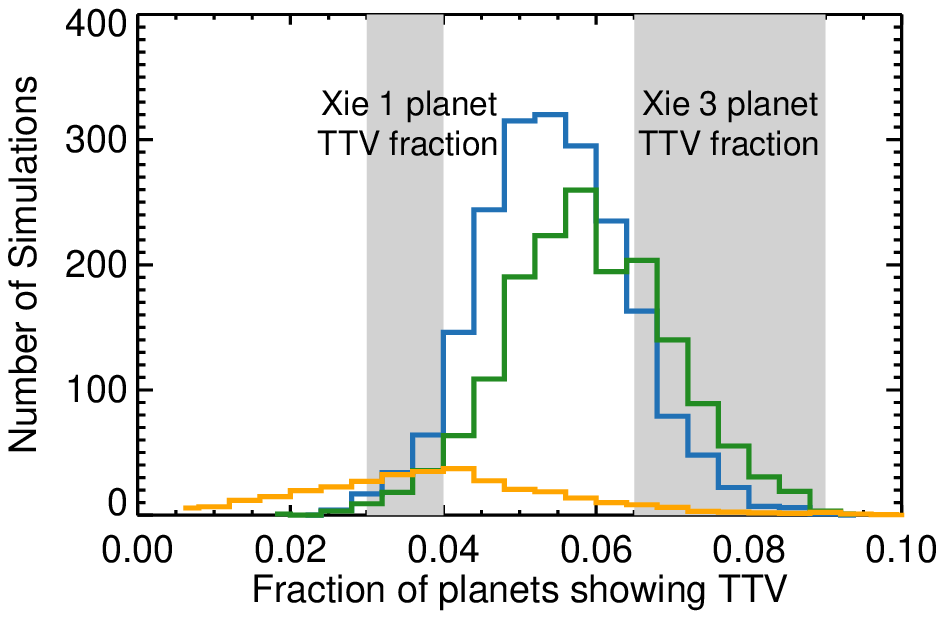} \\
  \end{tabular}
  \caption{{\it Top panel:} Posterior distribution for fraction of {\it Kepler} planets (black) exhibiting transit timing variations, as defined by ``Case 3" in \cite{Xie14}. The separate distributions for the dynamically cool (green) and dynamically hot (orange) populations have been scaled to reflect their relative contributions to the number of planets. {\it Bottom panel:} The same distributions predicted for the {\it TESS} sample.}
\label{fig:ttv}  
\end{figure}

While {\it TESS} itself may only rarely have the observational baseline to observe transit-timing variations, we can see predict their frequency among {\it TESS}-detected planets. Employing the empirical TTV likelihood as a function of number of transiting planets from \cite{Xie14}, we predict the TTV likelihood among the {\it TESS} transiting planets. Figure \ref{fig:ttv} shows a comparison of the TTV fraction for both missions. In the {\it Kepler} sample, the overall rate of 5\% reflects the mixture of planetary systems to which compact multis contribute only half. We show the TTV occurrence fraction from both populations in green (dynamically cool) and orange (dynamically hot), where the height of the histogram reflects the contribution of that population to the total number of planets. The {\it TESS} completeness, in contrast, heavily favors the types of compact multiples that exhibit TTV. For {\it TESS}, these types of planetary systems will comprise a likely 70\% of the yield, as we describe in the previous section. The fact that the final TTV rate is similar to {\it Kepler's} is due to a subtlety. Though compact multis are favored for detection by {\it TESS}, only one or two planets are typically detected in these systems, even if 3 or 4 transit. In comparison, consider the 3 and 4 transiting planets systems detected by {\it Kepler}. With the higher TTV probability per planet, the fact that there are 3 or 4 planets each with this higher probability (as opposed to 1 or 2) skews the overall TTV likelihood higher. The tradeoff between these two phenomena results in a TTV fraction similar to {\it Kepler}'s, despite {\it TESS}' strong preference for compact multiples in which TTV are more common. 

\subsection{Planet Bulk Density} 
The third panel of Figure \ref{fig:completeness2} shows the posterior distribution in theoretical density assigned from mutual Hill spacing per \cite{Dawson16}. Only planets in systems for which Hill spacing is applicable (those with with 2 or more planets) contribute to this distribution. We have indicated with cross-hatching the densities too high or low to be included in the \cite{Dawson16} metric. The higher fraction of compact multiples within {\it TESS}, and their accompanying close orbital spacing, maps to a predicted lower density on average. Drawing from the entire predicted {\it TESS} yield for M dwarf systems with 2 or more planets, we find $\bar{\rho}=0.3^{+0.3}_{-0.1}$. which are likely 70\% of the {\it TESS} planetary systems. The corresponding mean density for {\it Kepler} systems with 2 or more planets is $\bar{\rho}=0.9^{+0.6}_{-0.4}$. This prediction for the average fluffiness of {\it TESS} planets will ultimately be tested with radial velocity follow-up and transmission spectroscopy: we leave specific implications for those follow-up efforts for future work. 

\section{Summary and Conclusions}
\label{sec:conclusions}

Using the injected and detected samples published by \cite{Sullivan15}, we have extracted a completeness function with planet radius and orbital period for {\it TESS} M dwarfs. We first demonstrate that the application of this completeness function to the \cite{Sullivan15} injected planet sample correctly recovers the planet detections from that work. We then re-apply the completeness function, assuming a different planet occurrence rate. Rather than assuming 2--3 planets per star, we assume a mixture model with two types of planetary systems. One type contains $>5$ closely aligned planets (around 20\% of stars, per Section $\S$3.4), and one that contains one planet, or two planets with high mutual inclination respective to one another. We return to our enumerated list of goals from Section 1, to summarize our findings on each.  

 \begin{enumerate}
     \item We predict that {\it TESS} will uncover 990$\pm$350 planets orbiting 715$\pm$255 stars, a factor of 1.5 more than predicted in \cite{Sullivan15} ($\S$3.1). The error budget on number of detections is dominated by uncertainty on the underlying fraction of compact multiples in nature.
     \item Even given the typical duration of 27 days per star, we predict {\it TESS} will detect two or more planets around 140$\pm$70 stars among the hosts above. The approximate 20\% contribution of multis to the total host star budget is similar to {\it Kepler} M dwarfs. The high rate of compact multiples indicates that {\it TESS} will even detect three or more planets around 32$\pm$19 stars ($\S$3.2).
     \item Among 200 typical {\it TESS} M dwarf host stars, an average of 250 planets will be detectable in the mission data itself. We predict half that number lurk below the mission sensitivity: 116$^{+28}_{-28}$ planets per 200 host stars. Many of these planets will be readily detectable from ground-based surveys and space-based campaigns ($\S$3.3)
    \item We confirm the compact multiple rate (defined as 2 or more planets with orbital periods $<$10 days) among M dwarfs measured previously in the literature. We find this rate to be 15$\pm5$\% among early M dwarfs, as compared to 15.9$\pm$1.5\% \citep{Muirhead15}, using a different technique. While compact multiple hosts are not the majority in nature, the relative ease of their detection makes them overrepresented in transit surveys: they are 45$\pm$10\% of {\it Kepler}-detected planet hosts, and we predict they will be 68$\pm$12\% of {\it TESS}-detected planet hosts.
     \item By virtue of the lower average eccentricities of planets in multiple-planet systems, we predict that average orbital eccentricity of planets detected by {\it TESS} will correspondingly be lower than the average for {\it Kepler}. For systems in which {\it TESS} detects two or more transiting planets, 80\% of planets will have orbital eccentricities less than 0.1
     \item Despite the higher fraction of compact multiples in the {\it TESS} yield, the number of planets detected by {\it TESS} that exhibit transit timing variations (as defined by \citealt{Xie14}) will be similar to the overall 5\% observed for {\it Kepler}. These TTV will not be detectable by {\it TESS} itself as for {\it Kepler}, but we predict the underlying rate among detected planets to be similar. 
    \item Employing the planet formation theory of \cite{Dawson16} linking adjacent planet spacing to planet bulk density, we predict that {\it TESS} planets will be fluffier on average. We apply this metric to systems with at least two planets, to find $\bar{\rho}=0.3^{+0.3}_{-0.1}$ among {\it TESS} planets, as compared to $\bar{\rho}=0.9^{+0.6}_{-0.4}$ for {\it Kepler} planets.($\S$3.7)
 \end{enumerate}

We conclude by re-emphasizing the ground and space-based opportunity for photometric follow-up, specific to planet discovery. Around stars for which {\it TESS} detects one or more planets, we predict a wealth of additional transiting planets: present but undetectable in the {\it TESS} light curves (Section $\S$3.3). We take as an example a hypothetical survey of 200 {\it TESS} host stars, sensitive to 1 $R_{\oplus }$ planets and  with  100\%  completeness  out  to  only  2.2  days. Such an study will detect  11$\pm5$  additional  planets. Odds improve yet more if {\it TESS} has detected two or more planets around the star, or if one of the planets exhibits transit-timing variations. An extended {\it Spitzer} mission (as detailed by \citealt{Yee17}) would have the photometric sensitivity time baseline to be sensitive to rocky planets even in the habitable zone of {\it TESS} hosts. While we predict {\it TESS} itself to detect 2$^{+2}_{-1}$ planets <$1.25R_{\oplus}$ with orbit periods $20<P<40$ days, an additional 19$\pm$12 such planets will orbit known {\it TESS} hosts to another transiting planet, but elude detection by {\it TESS} proper. An extended {\it Spitzer} mission may be singularly suited for their discovery. 

The soon-to-be-launched CHaracterising ExOPlanet Satellite (CHEOPS, \citealt{Fortier14}) will gather high-precision transit observations to constrain (among many objectives) planetary atmospheres and formation. The telescope, in a geocentric orbit, will continuously point at a single target for a typical 6-12 hours; however, it can achieve stares of a few weeks' duration \citep{Broeg13}, and in principle could also readily detect additional planets. Additionally, ground-based photometric surveys such as MEarth \citep{Nutzman08} and TRAPPIST \citep{Gillon11} have already demonstrated the ability to detect planets $<2_{\oplus}$ orbiting M dwarfs \citep{Charbonneau09,Berta15,Dittmann17, Gillon17}. The forthcoming wealth of {\it TESS} planets portends a strong synergy with follow-up efforts. 

\section*{Acknowledgments}
We thank Philip Muirhead, Joshua Winn, and particularly John Johnson for discussions that immeasurably improved this manuscript. S.B. is funded by the MIT Torres Fellowship for Exoplanet Science.


\end{document}